\documentclass[conference,10pt]{IEEEtran}
\usepackage{mystyle}
\IEEEoverridecommandlockouts

\begin{document}

\title{Secure Degrees of Freedom of the Gaussian Wiretap Channel with Helpers and No Eavesdropper CSI: Blind Cooperative Jamming\thanks{This work was supported by NSF Grants CNS 09-64632, CCF 09-64645, CCF
10-18185 and CNS 11-47811.}}

\author{\authorblockN{Jianwei Xie \qquad \qquad Sennur Ulukus}
\authorblockA{Department of Electrical and Computer Engineering\\
University of Maryland, College Park, MD 20742\\
\emph{xiejw@umd.edu \qquad \qquad ulukus@umd.edu}}}

\maketitle
\thispagestyle{empty}
\pagestyle{empty}

\newcounter{small_constant}
\setcounter{small_constant}{0}
\newcommand{\nextsc}{\addtocounter{small_constant}{1}  c_{\arabic{small_constant}} }
\newcommand{\nextscnu}{c_{\arabic{small_constant}}}
\newcounter{small_constant_0}
\newcounter{small_constant_1}
\newcounter{small_constant_2}
\newcounter{small_constant_3}
\newcommand{\etX}{\bar{X}}

\maketitle

\begin{abstract}
We consider the Gaussian wiretap channel with $M$ helpers, where no eavesdropper channel state information (CSI) is available at the legitimate entities. The exact secure \dof of the Gaussian wiretap channel with $M$ helpers with perfect CSI at the transmitters was found in \cite{xie_gwch_allerton, xie_sdof_networks_in_prepare} to be $\frac{M}{M+1}$. One of the key ingredients of the optimal achievable scheme in \cite{xie_gwch_allerton, xie_sdof_networks_in_prepare} is to align cooperative jamming signals with the information symbols at the eavesdropper to limit the information leakage rate. This required perfect eavesdropper CSI at the transmitters. Motivated by the recent result in \cite{khisti_arti_noise_alignment}, we propose a new achievable scheme in which cooperative jamming signals span the \emph{entire space} of the eavesdropper, but are not exactly aligned with the information symbols. We show that this scheme achieves the same secure \dof of $\frac{M}{M+1}$ in \cite{xie_gwch_allerton, xie_sdof_networks_in_prepare} but does not require any eavesdropper CSI; the transmitters {\it blindly} cooperative jam the eavesdropper.
\end{abstract}

\section{Introduction}

Wyner introduced the wiretap channel in which a legitimate transmitter wants to have secure communications with a legitimate receiver in the presence of an eavesdropper, and determined its capacity-equivocation region for the degraded case \cite{wyner}. Csiszar and Korner extended this result to the general, not necessarily degraded, wiretap channel \cite{csiszar}. Leung-Yan-Cheong and Hellman determined the capacity-equivocation region of the Gaussian wiretap channel \cite{gaussian}. This line of research has been subsequently extended to many multi-user settings. Here, we are particularly interested in models with multiple independent legitimate transmitters, e.g., interference channel with confidential messages \cite{secrecy_ic, he_outerbound_gic_cm_ciss_09}, interference channel with external eavesdroppers \cite{koyluoglu_ic_external}, multiple access wiretap channel
\cite{tekin_gmac_w, cooperative_jamming, ersen_mac_allerton_08, liang_mac_cm_08,ersen_mac_book_chapter}, wiretap channel with helpers \cite{wiretap_channel_with_one_helper}, and relay-eavesdropper channel with deaf helpers \cite{relay_2}.

Since in most multi-user scenarios it is difficult to obtain the exact secrecy capacity region, recently, there has been a significant interest in studying the asymptotic performance of these systems at high signal-to-noise ratio (SNR) in terms of their secure degrees of freedom (d.o.f.) regions. Achievable secure \dof has been studied for several channel structures, such as the $K$-user Gaussian interference channel with confidential messages \cite{koyluoglu_k_user_gic_secrecy, he_k_gic_cm_09}, $K$-user interference channel with external eavesdroppers \cite{xie_k_user_ia_compound} in ergodic fading setting \cite{koyluoglu_k_user_gic_secrecy, koyluoglu_secrecy_scaling }, Gaussian wiretap channel with helpers \cite{secrecy_ia4, secrecy_ia_new, xiang_he_thesis, xie_gwch_allerton,xie_sdof_networks_in_prepare}, Gaussian multiple access wiretap channel \cite{secrecy_ia5} in ergodic fading setting \cite{raef_mac_it_12}, multiple antenna compound wiretap channel \cite{interference_alignment_compound_channel}, and wireless $X$ network \cite{secrecy_ia1}. The exact sum secure \dof was found for a large class of one-hop wireless networks, including the wiretap channel with $M$ helpers, two-user interference channel with confidential messages, and $K$-user multiple access wiretap channel in \cite{xie_sdof_networks_in_prepare}, and for all two-unicast layered wireless networks in \cite{xie_layered_network_isit_2012, xie_layered_network_journal}.

\begin{figure}[t]
\centering
\includegraphics[scale=0.7]{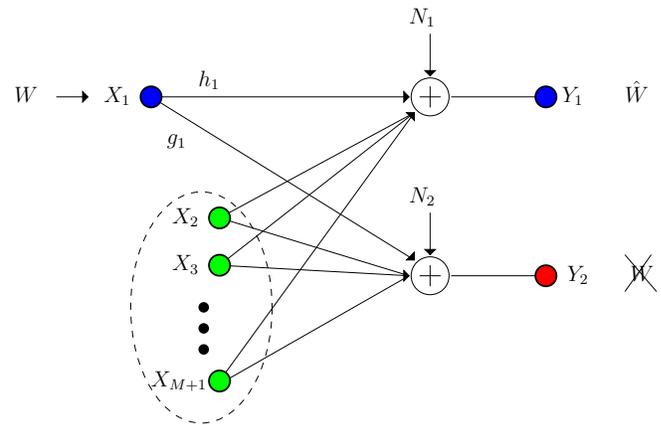}
\caption{The Gaussian wiretap channel with $M$ helpers.}
\label{fig:gwc_helper_general}
\end{figure}

\begin{figure*}[t]
\centering
\includegraphics[scale=0.8]{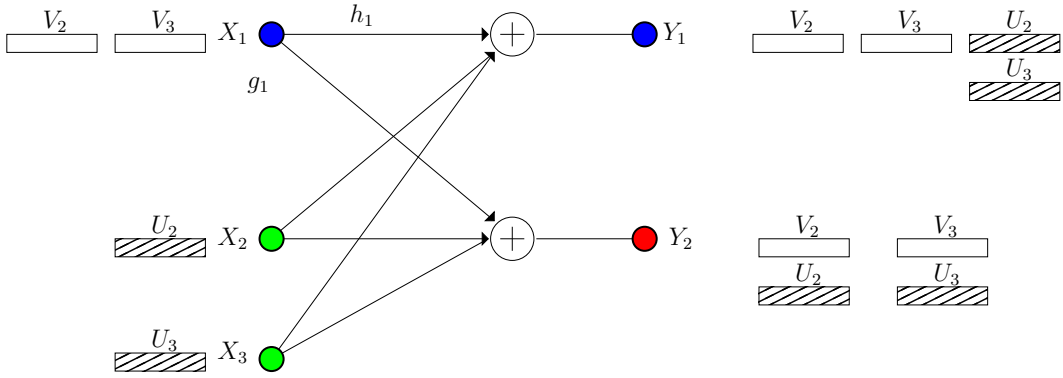}
\caption{Illustration of the alignment scheme for the Gaussian wiretap channel with $M$ helpers with eavesdropper's CSI available at all transmitters.}
\label{fig:gwc_no_csi_one_helper_ia}
\end{figure*}

In this paper, we revisit the Gaussian wiretap channel with $M$ helpers, see Fig.~\ref{fig:gwc_helper_general}. The secrecy capacity of the Gaussian wiretap channel with no helpers is the difference between the individual channel capacities of the transmitter-receiver and the
transmitter-eavesdropper pairs. This difference does not scale with the SNR, and hence the secure \dof of the Gaussian wiretap channel with no helpers is zero, indicating a severe penalty due to secrecy. It has been known that the secrecy rates can be improved if there are helpers which can transmit independent signals \cite{tekin_gmac_w, cooperative_jamming}, however, if the helpers transmit i.i.d. Gaussian signals, then the secure \dof is still zero \cite{raef_mac_it_12}. It has  been also known that positive secure \dof could be achieved if the helpers sent structured signals \cite{secrecy_ia4, secrecy_ia_new, xiang_he_thesis}, but the exact secure \dof was unknown. References \cite{xie_gwch_allerton, xie_sdof_networks_in_prepare} determined the exact secure \dof of the Gaussian wiretap channel with $M$ helpers to be $\frac{M}{M+1}$. This result was derived under the assumption that the eavesdropper's CSI was available at the transmitters. In the present paper, we show that the same secure \dof can be achieved even when the eavesdropper's CSI is unknown at the legitimate transmitters. This result is practically significant because, generally, it is difficult or impossible to obtain the eavesdropper's CSI. Since the upper bound developed in \cite{xie_gwch_allerton, xie_sdof_networks_in_prepare} is valid for this case also, we thus determine the exact secure \dof of the Gaussian wiretap channel with $M$ helpers with no eavesdropper CSI to be $\frac{M}{M+1}$. The achievable scheme in the case of no eavesdropper CSI here is significantly different than the achievable scheme with eavesdropper CSI developed in \cite{xie_gwch_allerton, xie_sdof_networks_in_prepare}.

In particular, in \cite{xie_gwch_allerton, xie_sdof_networks_in_prepare}, the legitimate transmitter divides its message into $M$ sub-messages and sends them on $M$ different {\it irrational dimensions}. Each one of the helpers sends a cooperative jamming signal. The message signals and the cooperative jamming signals are sent in such a way that: 1) the cooperative jamming signals are aligned at the legitimate receiver in the same irrational dimension, so that they occupy the smallest possible space at the legitimate receiver to enable the decodability of the message signals, and 2) each cooperative jamming signal is aligned exactly in the same irrational dimension with one of the message signals at the eavesdropper to protect it. This scheme is illustrated in Fig.~\ref{fig:gwc_no_csi_one_helper_ia} for $M=2$ helpers. In \cite{xie_gwch_allerton, xie_sdof_networks_in_prepare}, we used insights from \cite{secrecy_ia4,secrecy_ia_new, xiang_he_thesis} to show that, when a cooperative jamming signal is aligned with a message signal in the same irrational dimension at the eavesdropper, this alignment protects the message signal, and limits the information leakage rate to the eavesdropper by a constant which does not depend on the transmit power. Meanwhile, due to the alignment of the cooperative jamming signals in a small space at the legitimate receiver, the information rate to the legitimate receiver can be made to scale with the transmit power. We use this real interference alignment \cite{real_inter_align, real_inter_align_exploit} based approach to achieve a secure \dof of $\frac{M}{M+1}$ for {\it almost all channel gains}, and develop a converse to show that it is in fact the secure \dof capacity.

\begin{figure*}[t]
\centering
\includegraphics[scale=0.8]{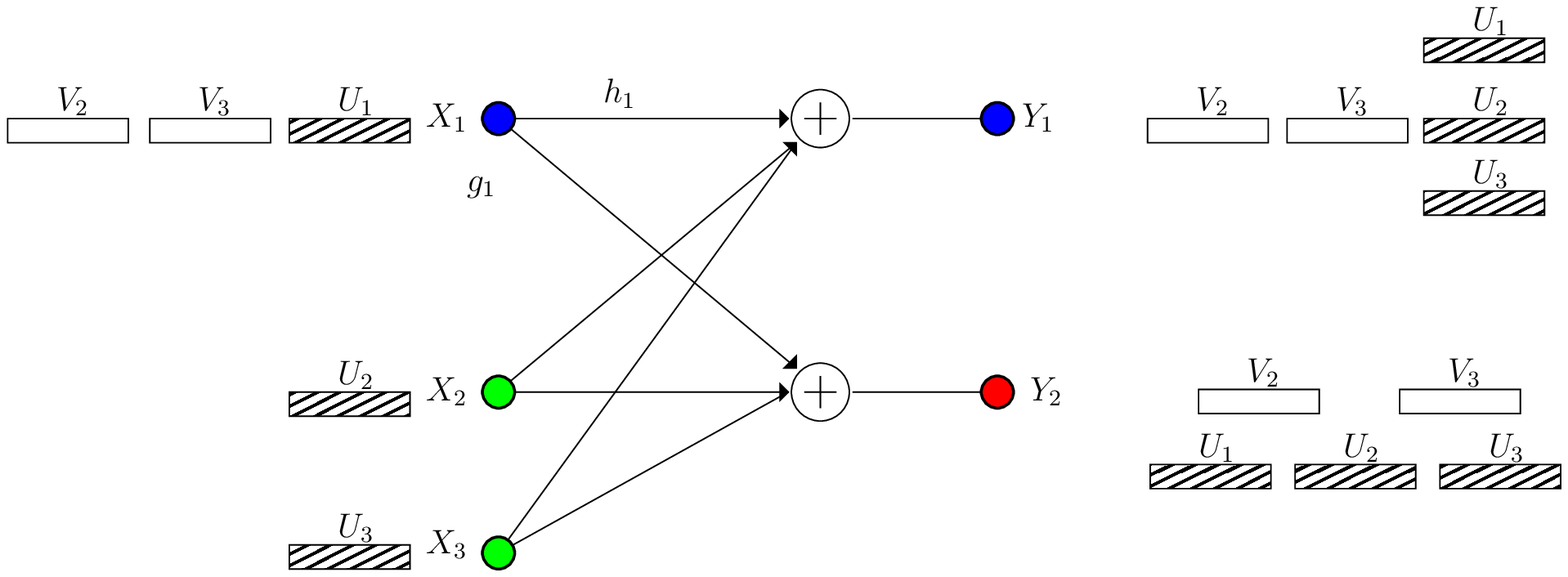}
\caption{Illustration of the alignment scheme for the Gaussian wiretap channel with $M$ helpers with no eavesdropper CSI.}
\label{fig:gwc_no_csi_one_helper_no_csi_ia}
\vspace{-0.05in}
\end{figure*}

The achievable scheme in the present paper again divides the message into $M$ sub-messages. Each one of the helpers sends a cooperative jamming signal. As a major difference from the achievable scheme in \cite{xie_gwch_allerton, xie_sdof_networks_in_prepare}, in this achievable scheme, the legitimate transmitter also sends a cooperative jamming signal. This scheme is illustrated in Fig.~\ref{fig:gwc_no_csi_one_helper_no_csi_ia} for $M=2$ helpers. In this case, the message signals and the cooperative jamming signals are sent in such a way that: 1) all $M+1$ cooperative jamming signals are aligned at the legitimate receiver in the same irrational dimension, and 2) all cooperative jamming signals span the {\it entire space} at the eavesdropper to limit the information leakage to the eavesdropper. We use insights from \cite{khisti_arti_noise_alignment}, which developed a new achievable scheme that achieved the same secure \dof as in \cite{interference_alignment_compound_channel} without eavesdropper CSI, to show that the information leakage to the eavesdropper is upper bounded by a function, which can be made arbitrarily small. On the other hand, since the cooperative jamming signals occupy the smallest space at the legitimate receiver, the information rate to the legitimate receiver can be made to scale with the transmit power. In this achievable scheme, we let the legitimate transmitter and the helpers {\it blindly} cooperative jam the eavesdropper. Because of the inefficiency of {\it blind} cooperative jamming, here, we had to use more cooperative jamming signals than in \cite{xie_gwch_allerton, xie_sdof_networks_in_prepare}, i.e., in \cite{xie_gwch_allerton, xie_sdof_networks_in_prepare} we use a total of $M$ cooperative jamming signals from the helpers, while here we use $M+1$ cooperative jamming signals, one of which coming from the legitimate transmitter.

\section{System Model and Definitions}

The Gaussian wiretap channel with $M$ helpers, see Fig.~\ref{fig:gwc_helper_general}, is defined by
\begin{align}
\label{eqn:gwch_channel_model_helpers_genneral_1}
Y_1 &= h_1 X_1 + \sum_{j=2}^{M+1}h_j X_j + N_1 \\
\label{eqn:gwch_channel_model_helpers_genneral_2}
Y_2 & = g_1 X_1 + \sum_{j=2}^{M+1}g_j X_j + N_2
\end{align}
where $Y_1$ is the channel output of the legitimate receiver, $Y_2$ is the channel output of the eavesdropper, $X_1$ is the channel input of the legitimate transmitter, $X_i$, for $i=2,\ldots,M+1$, are the channel inputs of the $M$ helpers, $h_i$ is the channel gain of the $i$th transmitter to the legitimate receiver, $g_i$ is the channel gain of the $i$th transmitter to the eavesdropper, and $N_1$ and $N_2$ are two independent zero-mean unit-variance Gaussian random variables. All channel inputs satisfy average power constraints, $\E\left[X^2_{i}\right] \le P$, for $i=1,\ldots, M+1$.

Transmitter $1$ intends to send a message $W$, uniformly chosen from a set $\mathcal{W}$, to the legitimate receiver (receiver $1$). The rate of the message is $R\defn\frac{1}{n}\log|\mathcal{W}|$, where $n$ is the number of channel uses. Transmitter $1$ uses a stochastic function $f: \mathcal{W}\to
\bfX_1$ to encode the message, where $\bfX_1\defn X_1^n$ is the $n$-length channel input. We use boldface letters to denote $n$-length vector signals, e.g., $\bfX_1\defn X_1^n$, $\bfY_1\defn Y_1^n$, $\bfY_2\defn Y_2^n$, etc. The legitimate receiver decodes the message as $\hat{W}$ based on its observation $\mathbf{Y}_1$. A secrecy rate $R$ is said to be achievable if for any $\epsilon>0$ there
exists an $n$-length code such that receiver $1$ can decode this message reliably, i.e., the probability of decoding error is less than $\epsilon$,
\begin{equation}
\label{eqn:reliability_measure}
\pr\left[W\neq\hat{W}\right] \le \epsilon
\end{equation}
and the message is kept information-theoretically secure against the eavesdropper,
\begin{equation}
\label{eqn:secrecy_measure}
\frac{1}{n}H(W| \bfY_2) \ge \frac{1}{n} H(W) - \epsilon
\end{equation}
i.e., that the uncertainty of the message $W$, given the observation $\bfY_2$ of the eavesdropper, is almost equal to the entropy of the message. The supremum of all achievable secrecy rates is the secrecy capacity $C_s$, and the secure d.o.f., $D_s$, is defined as
\begin{equation}
D_s \defn \lim_{P\to\infty}  \frac{C_s}{\frac{1}{2}\log P}
\label{eqn:sec-dof-defn}
\end{equation}

Note that $D_s\le 1$ is an upper bound. To avoid trivial cases, we assume that $h_1\neq 0$ and $g_1\neq 0$. Without the independent helpers, i.e., $M=0$, and with full knowledge of all channel gains,  the secrecy capacity of the Gaussian wiretap channel is known \cite{gaussian}
\begin{equation}
C_s = \frac{1}{2}\log\left(1+h_1^2 P\right) -  \frac{1}{2}\log\left(1+g_1^2 P\right)
\end{equation}
and from \eqn{eqn:sec-dof-defn} the secure \dof is zero. Therefore, we assume $M\ge 1$. If there exists a $j$ ($j=2,\ldots,M+1$) such that $h_j=0$ and $g_j\neq 0$, then a lower bound of $1$ secure \dof  can be obtained for this channel by letting this helper jam the eavesdropper by i.i.d.~Gaussian noise of power $P$ and keeping all other helpers silent. This lower bound matches the upper bound, giving the secure \dof On the other hand, if there exists a $j$ ($j=2,\ldots,M+1$) such that $h_j\neq0$ and $g_j= 0$, then this helper can be removed from the channel model without affecting the secure d.o.f. Therefore, in the rest of the paper, we  assume that $M\ge1$ and $h_j\neq0$ and $g_j\neq 0$
for all $j=1,\cdots,M+1$.

\section{Achievable Scheme with no Eavesdropper CSI}

In this section, we propose an achievable scheme to achieve the secure \dof of $\frac{M}{M+1}$ with no eavesdropper CSI at any of the transmitters. The only assumption we make is that the legitimate transmitter knows an upper bound of $\sum_{k=1}^{M+1}g_k^2\leq \bar{c}$ on the eavesdropper channel gains.

Let $\{V_2,V_3,\cdots,V_{M+1},U_1,U_2,U_3,\cdots,U_{M+1}\}$ be mutually independent discrete random variables, each of which  uniformly drawn from the same PAM constellation $C(a,Q)$
\begin{equation}
C(a,Q) = a \{ -Q, -Q+1, \ldots, Q-1,Q\}
\end{equation}
where $Q$ is a positive integer and $a$ is a real number used to normalize the transmission power, and is also the minimum distance between the points belonging to $C(a,Q)$. Exact values of $a$ and $Q$ will be specified later. We choose the input signal of the legitimate transmitter as
\begin{equation}
X_1  = \frac{1}{h_1}U_1 +  \sum_{k=2}^{M+1} \alpha_k  V_k
\label{legit-signal}
\end{equation}
where $\{\alpha_k\}^{M+1}_{k=2}$ are rationally independent among themselves and also rationally independent of all channel gains. The input signal of the $j$th helper, $j=2,3,\cdots,M+1$, is chosen as
\begin{equation}
X_j = \frac{1}{h_j} U_j
\label{helper-signal}
\end{equation}
Note that, neither the legitimate transmitter signal in (\ref{legit-signal}) nor the helper signals in (\ref{helper-signal}) depend on the eavesdropper CSI $\{g_k\}_{k=1}^{M+1}$. With these selections, observations of the receivers are given by,
\begin{align}
Y_1 & = \sum_{k=2}^{M+1} {h_1 \alpha_k} V_k + \left( \sum_{j=1}^{M+1} U_j \right)+ N_1 \\
Y_2 & = \sum_{k=2}^{M+1} {g_1 \alpha_k} V_k +  \sum_{j=1}^{M+1} \frac{g_j}{h_j} U_j + N_2
\end{align}

The intuition here is as follows: We use $M$ independent sub-signals $V_k$, $k=2,3,\cdots,M+1$, to represent the original message $W$. The input signal $X_1$ is a linear combination of $V_k$s and a jamming signal $U_1$. At the legitimate receiver, all of the cooperative jamming signals, $U_k$s, are
aligned such that they occupy a small portion of the signal space. Since $\left\{1, h_1 \alpha_2, h_1 \alpha_3, \cdots, h_1 \alpha_{M+1}\right\}$ are rationally independent for  all channel gains, except for a set of Lebesgue measure zero, the signals $\left\{V_2,V_3,\cdots,V_{M+1}, \sum_{j=1}^{M+1} U_j \right\}$ can be distinguished by the legitimate receiver. In addition, we observe that $\left\{\frac{g_1}{h_1}, \cdots, \frac{g_{M+1}}{h_{M+1}}\right\}$ are rationally independent, and therefore, $\left\{U_1,U_2,\cdots,U_{M+1}\right\}$ \emph{span} the \emph{entire space} at the eavesdropper; see Fig.~\ref{fig:gwc_no_csi_one_helper_no_csi_ia}.  Here, by the \emph{entire space}, we mean the maximum number of \emph{dimensions} that the eavesdropper is capable of decoding, which is $M+1$ in this case. Since the \emph{entire space} at the eavesdropper is occupied by the cooperative jamming signals, the message signals $\{V_2, V_3, \cdots, V_{M+1}\}$
are secure, as we will mathematically prove in the sequel.

Since, for $j\neq 1$, ${\bfX}_j$ is an i.i.d.~sequence and is independent of ${\bfX}_1$, the following secrecy rate is achievable \cite{csiszar}
\begin{equation}
C_s \ge I(\bfV;Y_1) - I(\bfV;Y_2)
\label{eqn:gwcnc-lower-bound}
\end{equation}
where $\bfV \defn \{V_2,V_3,\cdots, V_{M+1}\}$.

First, we  use Fano's inequality to bound the first term in \eqn{eqn:gwcnc-lower-bound}. Note that the \emph{space} observed at receiver $1$ consists of $(2Q+1)^M (2MQ+2Q+1)$ points in $M+1$ \emph{dimensions}, and the sub-signal in each \emph{dimension} is drawn from a constellation of $C(a,(M+1)Q)$. Here, we use the property that $C(a,Q)\subset C(a,(M+1)Q)$.  By using the Khintchine-Groshev theorem of Diophantine approximation in number theory \cite{real_inter_align, real_inter_align_exploit}, we can bound the minimum distance $d_{min}$ between the points
in receiver 1's \emph{space} as follows: For any $\delta>0$, there exists a constant $k_\delta$ such that
\begin{equation}
\label{eqn:gwch_lb_of_d_m_helper}
d_{min} \ge \frac{ k_\delta  a}{((M+1)Q)^{M+\delta}}
\end{equation}
for almost all rationally independent $\left\{1, h_1 \alpha_2, h_1 \alpha_3, \cdots, h_1 \alpha_{M+1}\right\}$, except for a set of Lebesgue measure zero. Then, we can upper bound the
probability of decoding error of such a PAM scheme by considering the additive
Gaussian noise at receiver $1$,
\begin{align}
\pe\left[\bfV\neq\hat{\bfV} \right]
& \le \exp\left(  - \frac{d_{min}^2}{8}\right)   \\
& \le \exp\left(  -
\frac{a^2k_\delta^2}{8 ((M+1)Q)^{2(M+\delta)}}\right)
\end{align}
where $\hat{\bfV}$ is the estimate of $\bfV$ by choosing the closest point in the constellation based on observation $Y_1$. For any $\delta>0$, if we choose $Q = P^{\frac{1-\delta}{2(M+1+\delta)}}$ and $a=\gamma P^{\frac{1}{2}}/Q$, where  $\gamma$
is a constant independent of $P$, then
\begin{align}
\pe\left[\bfV\neq\hat{\bfV}\right]
& \le \exp\left( -\frac{k_\delta^2 \gamma^2 (M+1)^2 P}{8 ((M+1)Q)^{2(M+\delta)+2}} \right)
\\ &
 = \exp\left( -\frac{k_\delta^2 \gamma^2 (M+1)^2 P^\delta}{8 (M+1)^{2(M+1+\delta)}} \right)
\end{align}
and we can have $\pe\left[\bfV\neq\hat{\bfV}\right] \to 0$  as $P\to\infty$. To
satisfy the power constraint at the transmitters, we can simply choose
\begin{align}
& \gamma  \le \min\left\{\left[\frac{1}{|h_1| } + \sum_{k=2}^{M+1} |\alpha_k|\right]^{-1},|h_2|, |h_3|, \cdots,
|h_{M+1}|\right\}
\end{align}
By Fano's inequality and the Markov chain $\bfV\rightarrow
Y_1\rightarrow\hat{\bfV}$, we know that
\begin{align}
&H(\bfV | Y_1) \nl
&\quad \le H(\bfV|\hat{\bfV}) \\
&\quad \le 1 +
\exp\left( -\frac{k_\delta^2 \gamma^2 (M+1)^2 P^\delta}{8 (M+1)^{2(M+1+\delta)}}
\right) \log(2Q+1)^M
\\
& \quad = o(\log P)
\label{eqn:nocsi_fano_logp}
\end{align}
where $\delta$ and $\gamma$  are fixed, and  $o(\cdot)$ is the little-$o$ function. This means that
\begin{align}
  I(\bfV;Y_1)
& = H(\bfV) - H(\bfV|Y_1) \\
& \ge
 H(\bfV) - o(\log P) \\
& =
 \log(2Q+1)^M - o(\log P) \\
& \ge  \log P^{\frac{M(1-\delta)}{2(M+1+\delta)}} - o(\log P) \\
& =  {\frac{M(1-\delta)}{M+1+\delta}} \left( \frac{1}{2} \log P \right) - o(\log P) 
 \label{eqn:gwch_wiretap_m_helper_lb_ixy1}
\end{align}

Next, we need to bound the second term in \eqn{eqn:gwcnc-lower-bound},
\begin{align}
I(\bfV;Y_2)
& = I(\bfV,\bfU;{Y}_2) - I(\bfU;{Y}_2|\bfV) \\
& = I(\bfV,\bfU;{Y}_2) - H(\bfU|\bfV) +  H(\bfU|{Y_2},\bfV) \\
& = I(\bfV,\bfU;{Y}_2) - H(\bfU) +  H(\bfU|{Y_2},\bfV)
\\
& = h({Y}_2) - h({Y}_2|\bfV,\bfU) \nl
& \quad  - H(\bfU) +  H(\bfU|{Y_2},\bfV)
\\
& = h({Y}_2) - h(N_2)- H(\bfU) +  H(\bfU|{Y_2},\bfV)
\\
& \le h({Y}_2) - h(N_2)- H(\bfU) +  o(\log P)
\label{eqn:gwch_nocsi_decode_U_given_Y_and_V}
\\
& \le \frac{1}{2} \log 2\pi e( 1 + \bar{c} P) - \frac{1}{2} \log 2 \pi e \nl
& \quad  - \log(2Q+1)^{M+1} +  o(\log P) \\
& \le \frac{1}{2} \log  P  - \frac{(M+1)(1-\delta)}{2(M+1+\delta)}\log P +
o(\log P) \\
& =  \frac{(M+2)\delta}{M+1+\delta}\left( \frac{1}{2}\log P \right) +
o(\log P)
\label{eqn:gwch_wiretap_m_helper_lb_ixy2}
\end{align}
where $\bfU\defn\{U_1,U_2,\cdots,U_{M+1}\}$ and $\bar{c}$ is the upper bound on $\sum_{k=1}^{M+1}g_k^2$ defined at the beginning of this section, and \eqn{eqn:gwch_nocsi_decode_U_given_Y_and_V} is due to the  fact that given $\bfV$ and ${Y}_2$, the eavesdropper can decode $\bfU$ with probability of error
approaching zero since $\left\{\frac{g_1}{h_1}, \cdots, \frac{g_{M+1}}{h_{M+1}}\right\}$ are rationally independent for all channel gains, except for a set of Lebesgue measure zero. Then, by Fano's inequality, $H(\bfU|Y_2,\bfV)\le o(\log P)$ similar to the step in \eqn{eqn:nocsi_fano_logp}.

Combining \eqn{eqn:gwch_wiretap_m_helper_lb_ixy1} and \eqn{eqn:gwch_wiretap_m_helper_lb_ixy2}, we have
\begin{align}
C_s
& \ge I(\bfV;Y_1) - I(\bfV;Y_2) \\
& \ge  {\frac{M(1-\delta)}{M+1+\delta}} \left( \frac{1}{2} \log P \right) \nl
& \quad   - \frac{(M+2)\delta}{M+1+\delta}\left( \frac{1}{2}\log P \right) -
o(\log P)
 \\
& ={\frac{M - (2M+2)\delta}{M+1+\delta}} \left(\frac{1}{2}\log P\right) - o(\log P)
\end{align}
where again $o(\cdot)$ is the little-$o$ function. If we choose $\delta$ arbitrarily small, then we can achieve $\frac{M}{M+1}$ secure \dof for this model where there is no eavesdropper CSI at the transmitters.

\section{Conclusions}

We studied the Gaussian wiretap channel with $M$ helpers without any eavesdropper CSI at the transmitters. We proposed an achievable scheme that achieves a secure \dof of $\frac{M}{M+1}$, which is the same as the secure \dof reported in \cite{xie_gwch_allerton, xie_sdof_networks_in_prepare} when the transmitters had perfect eavesdropper CSI. The new achievability scheme is based on real interference alignment and {\it blind} cooperative jamming. While \cite{xie_gwch_allerton, xie_sdof_networks_in_prepare} aligned cooperative jamming signals with the information symbols at the eavesdropper to protect the information symbols, which required eavesdropper CSI, here we used one more cooperative jamming signal to span the {\it entire space} at the eavesdropper to protect the information symbols. As in \cite{xie_gwch_allerton, xie_sdof_networks_in_prepare}, here also, we aligned all of the cooperative jamming signals in the same dimension at the legitimate receiver, in order to occupy the smallest space at the legitimate receiver to allow for the decodability of the information symbols. Therefore, we aligned the cooperative jamming signals carefully only at the legitimate receiver, which required only the legitimate receiver's CSI at the transmitters.

\end{document}